
\endcomment
%
\input amstex
%
%
%
%


\magnification=1200
\hsize=31pc
\vsize=55 truepc
\hfuzz=2pt
\vfuzz=4pt
\pretolerance=500
\tolerance=500
\parskip=0pt plus 1pt
\parindent=16pt
%

%
%
\font\fourteenrm=cmr10 scaled \magstep2
\font\fourteeni=cmmi10 scaled \magstep2
\font\fourteenbf=cmbx10 scaled \magstep2
\font\fourteenit=cmti10 scaled \magstep2
\font\fourteensy=cmsy10 scaled \magstep2

%
\font\large=cmbx10 scaled \magstep1

%
\font\sans=cmssbx10

%

%

%
\font\eightrm=cmr8
\font\eighti=cmmi8
\font\eightbf=cmbx8
\font\eightit=cmti8

\font\eightsy=cmsy8
\font\sixrm=cmr6
\font\sixi=cmmi6
\font\sixsy=cmsy6

%
\def\tenpoint{\def\rm{\fam0\tenrm}%
  \textfont0=\tenrm \scriptfont0=\sevenrm
                      \scriptscriptfont0=\fiverm
  \textfont1=\teni  \scriptfont1=\seveni
                      \scriptscriptfont1=\fivei
  \textfont2=\tensy \scriptfont2=\sevensy
                      \scriptscriptfont2=\fivesy
  \textfont3=\tenex   \scriptfont3=\tenex
                      \scriptscriptfont3=\tenex
  \textfont\itfam=\tenit  \def\it{\fam\itfam\tenit}%
  \textfont\slfam=\tensl  \def\sl{\fam\slfam\tensl}%
  \textfont\bffam=\tenbf  \scriptfont\bffam=\sevenbf
                            \scriptscriptfont\bffam=\fivebf
                            \def\bf{\fam\bffam\tenbf}%
  \normalbaselineskip=20 truept
  \setbox\strutbox=\hbox{\vrule height14pt depth6pt
width0pt}%
  \let\sc=\eightrm \normalbaselines\rm}
\def\eightpoint{\def\rm{\fam0\eightrm}%
  \textfont0=\eightrm \scriptfont0=\sixrm
                      \scriptscriptfont0=\fiverm
  \textfont1=\eighti  \scriptfont1=\sixi
                      \scriptscriptfont1=\fivei
  \textfont2=\eightsy \scriptfont2=\sixsy
                      \scriptscriptfont2=\fivesy
  \textfont3=\tenex   \scriptfont3=\tenex
                      \scriptscriptfont3=\tenex
  \textfont\itfam=\eightit  \def\it{\fam\itfam\eightit}%
  \textfont\bffam=\eightbf  \def\bf{\fam\bffam\eightbf}%
  \normalbaselineskip=16 truept
  \setbox\strutbox=\hbox{\vrule height11pt depth5pt width0pt}}
\def\fourteenpoint{\def\rm{\fam0\fourteenrm}%
  \textfont0=\fourteenrm \scriptfont0=\tenrm
                      \scriptscriptfont0=\eightrm
  \textfont1=\fourteeni  \scriptfont1=\teni
                      \scriptscriptfont1=\eighti
  \textfont2=\fourteensy \scriptfont2=\tensy
                      \scriptscriptfont2=\eightsy
  \textfont3=\tenex   \scriptfont3=\tenex
                      \scriptscriptfont3=\tenex
  \textfont\itfam=\fourteenit  \def\it{\fam\itfam\fourteenit}%
  \textfont\bffam=\fourteenbf  \scriptfont\bffam=\tenbf
                             \scriptscriptfont\bffam=\eightbf
                             \def\bf{\fam\bffam\fourteenbf}%
  \normalbaselineskip=24 truept
  \setbox\strutbox=\hbox{\vrule height17pt depth7pt width0pt}%
  \let\sc=\tenrm \normalbaselines\rm}
\def\today{\number\day\ \ifcase\month\or
  January\or February\or March\or April\or May\or June\or
  July\or August\or September\or October\or November\or
December\fi
  \space \number\year}
\def\monthyear{\ifcase\month\or
  January\or February\or March\or April\or May\or June\or
  July\or August\or September\or October\or November\or
December\fi
  \space \number\year}

%
\newcount\secno      
\newcount\subno      
\newcount\subsubno   
\newcount\appno      
\newcount\tableno    
\newcount\figureno   
%

%
\normalbaselineskip=20 truept
\baselineskip=20 truept

%
%
\def\title#1
   {\vglue1truein
   {\baselineskip=24 truept
    \pretolerance=10000
    \raggedright
    \noindent \fourteenpoint\bf #1\par}
    \vskip1truein minus36pt}
%

%
\def\author#1
  {{\pretolerance=10000
    \raggedright
    \noindent {\large #1}\par}}

%
\def\address#1
   {\bigskip
    \noindent \rm #1\par}

%
\def\shorttitle#1
   {\vfill
    \noindent \rm Short title: {\sl #1}\par
    \medskip}

%
\def\pacs#1
   {\noindent \rm PACS number(s): #1\par
    \medskip}

%
\def\jnl#1
   {\noindent \rm Submitted to: {\sl #1}\par
    \medskip}

%
\def\date
   {\noindent Date: \today\par
    \medskip}

%

%
\def\keyword#1
   {\bigskip
    \noindent {\bf Keyword abstract: }\rm#1}

%

%
%

%
\def\entry#1#2#3
   {\noindent
    \hangindent=20pt
    \hangafter=1
    \hbox to20pt{#1 \hss}#2\hfill #3\par}

%
\def\subentry#1#2#3
   {\noindent
    \hangindent=40pt
    \hangafter=1
    \hskip20pt\hbox to20pt{#1 \hss}#2\hfill #3\par}
\def\checkforsub{\futurelet\nexttok\decide}
\def\ssf{\relax}
\def\decide{\if\nexttok\ssf\let\endspace=\nospace
                \else\let\endspace=\extraspace\fi\endspace}
\def\nospace{\nobreak\par\nobreak}
%
%
\def\section#1{%
    \goodbreak
    \vskip24pt plus12pt minus12pt
    \nobreak
    \gdef\extraspace{\nobreak\bigskip\noindent\ignorespaces}%
    \noindent
    \subno=0 \subsubno=0
    \global\advance\secno by 1
    \noindent {\bf \the\secno. #1}\par\checkforsub}

%
\def\subsection#1{%
     \goodbreak
     \vskip24pt plus12pt minus6pt
     \nobreak
     \gdef\extraspace{\nobreak\medskip\noindent\ignorespaces}%
     \noindent
     \subsubno=0
     \global\advance\subno by 1
     \noindent {\sl \the\secno.\the\subno. #1\par}\checkforsub}

%
\def\subsubsection#1{%
     \goodbreak
     \vskip15pt plus6pt minus6pt
     \nobreak\noindent
     \global\advance\subsubno by 1
     \noindent {\sl \the\secno.\the\subno.\the\subsubno. #1}\null.
     \ignorespaces}

%
\def\appendix#1
   {\vskip0pt plus.1\vsize\penalty-250
    \vskip0pt plus-.1\vsize\vskip24pt plus12pt minus6pt
    \subno=0
    \global\advance\appno by 1
    \noindent {\bf Appendix \the\appno. #1\par}
    \bigskip
    \noindent}

%
\def\subappendix#1
   {\vskip-\lastskip
    \vskip36pt plus12pt minus12pt
    \bigbreak
    \global\advance\subno by 1
    \noindent {\sl \the\appno.\the\subno. #1\par}
    \nobreak
    \medskip
    \noindent}

%


%

%
\def\tabcaption#1
   {\global\advance\tableno by 1
    \noindent {\bf Table \the\tableno.} \rm#1\par
    \bigskip}

%

%

%

%

%

%
\def\figcaption#1
   {\global\advance\figureno by 1
    \noindent {\bf Figure \the\figureno.} \rm#1\par
    \bigskip}

%

%

%
\def\refjl#1#2#3#4
   {\hangindent=16pt
    \hangafter=1
    \rm #1
   {\frenchspacing\sl #2
    \bf #3}
    #4\par}

%
\def\refbk#1#2#3
   {\hangindent=16pt
    \hangafter=1
    \rm #1
   {\frenchspacing\sl #2}
    #3\par}

%
\def\numrefjl#1#2#3#4#5
   {\parindent=40pt
    \hang
    \noindent
    \rm {\hbox to 30truept{\hss #1\quad}}#2
   {\frenchspacing\sl #3\/
    \bf #4}
    #5\par\parindent=16pt}

%
\def\numrefbk#1#2#3#4
   {\parindent=40pt
    \hang
    \noindent
    \rm {\hbox to 30truept{\hss #1\quad}}#2
   {\frenchspacing\sl #3\/}
    #4\par\parindent=16pt}

%

\def\ref#1{\par\noindent \hbox to 21pt{\hss
#1\quad}\frenchspacing\ignorespaces}

%
\def\frac#1#2{{#1 \over #2}}

%

%

%
\def\e{\operatorname{e}}


\def\i{\operatorname{i}}
\chardef\ii="10

%

%

%

\catcode`\@=11
\def\vfootnote#1{\insert\footins\bgroup
    \interlinepenalty=\interfootnotelinepenalty
    \splittopskip=\ht\strutbox 
    \splitmaxdepth=\dp\strutbox \floatingpenalty=20000
    \leftskip=0pt \rightskip=0pt \spaceskip=0pt \xspaceskip=0pt
    \noindent\eightpoint\rm #1\ \ignorespaces\footstrut\futurelet\next\fo@t}

%
%
\def\eq(#1){\hfill\llap{(#1)}}
\catcode`\@=12
%
%



%
%





%
%

%
%

%
%

%

%

%

%
\def\gap{\;\lower3pt\hbox{$\buildrel > \over \sim$}\;}
%
%
\def\lap{\;\lower3pt\hbox{$\buildrel < \over \sim$}\;}
\def\tqs{\hbox to 25pt{\hfil}}



{\obeylines\gdef\startdisplay#1
  {\catcode`\^^M=5$$#1\halign\bgroup\indent##\hfil&&\qquad##\hfil\cr}}
\outer\def\enddisplay{\crcr\egroup$$}

\chardef\other=12
\def\ttverbatim{\begingroup \catcode`\\=\other \catcode`\{=\other
  \catcode`\}=\other \catcode`\$=\other \catcode`\&=\other
  \catcode`\#=\other \catcode`\%=\other \catcode`\~=\other
  \catcode`\_=\other \catcode`\^=\other
  \obeyspaces \obeylines \tt}
{\obeyspaces\gdef {\ }}  

\outer\def\begintt{$$\let\par=\endgraf \ttverbatim \parskip=0pt
  \catcode`\|=0 \rightskip=-5pc \ttfinish}
{\catcode`\|=0 |catcode`|\=\other 
  |obeylines 
  |gdef|ttfinish#1^^M#2\endtt{#1|vbox{#2}|endgroup$$}}

\catcode`\|=\active
{\obeylines\gdef|{\ttverbatim\spaceskip=.5em plus.25em minus.15em
                                            \let^^M=\ \let|=\endgroup}}%

\TagsOnRight
\tracingstats=1    

\font\twelverm=cmr10 scaled 1200

\def\CD{{\Cal D}}

\def\GV{G^{(V)}}

\def\GWa{G^{(Wall)}}

\def\Gd{G^{(\delta)}}
\def\Kd{K^{(\delta)}}

\def\ih{{\i\over\hbar}}
\def\bbbr{\operatorname{{I\!R}}}                     

\font\sans=cmssbx10
\def\sf{\sans}
\def\bbbz{{\mathchoice {\hbox{$\sf\textstyle Z\kern-0.4em Z$}}
{\hbox{$\sf\textstyle Z\kern-0.4em Z$}}
{\hbox{$\sf\scriptstyle Z\kern-0.3em Z$}}
{\hbox{$\sf\scriptscriptstyle Z\kern-0.2em Z$}}}}    

\def\arctan{\operatorname{arctan}}
\def\myalign{\allowdisplaybreaks\align}
\def\dfrac{\dsize\frac}

\hfuzz=3pt

\newcount\glno
\def\plus{\advance\glno by 1}
\def\num{\the\glno}

\newcount\Refno
\def\add{\advance\Refno by 1}
\Refno=1

\edef\BADU{\the\Refno}\add
\edef\CMS{\the\Refno}\add
\edef\CAR{\the\Refno}\add
\edef\DK{\the\Refno}\add
\edef\GRSb{\the\Refno}\add
\edef\KLE{\the\Refno}\add
\edef\FH{\the\Refno}\add
\edef\GODE{\the\Refno}\add
\edef\AGHH{\the\Refno}\add
\edef\GROh{\the\Refno}\add
\edef\GROe{\the\Refno}\add
\edef\KLEMUS{\the\Refno}\add
\edef\FLU{\the\Refno}\add


{\nopagenumbers
\pageno=0
\centerline{\hfill hep-th/9303128}
\smallskip
\centerline{\monthyear\hfill SISSA/46/93/FM}
\vskip1cm
\centerline{\fourteenbf
     Path Integration Via Summation of Perturbation}
\centerline{\fourteenbf
     Expansions and Applications to Totally Reflecting}
\centerline{\fourteenbf
     Boundaries, and Potential Steps}
\vskip1cm
\centerline{\twelverm CHRISTIAN GROSCHE}
\bigskip
\centerline{\it Scuola Internazionale Superiore di Studi Avanzati}
\centerline{\it International School for Advanced Studies}
\centerline{\it Via Beirut 4, 34014 Trieste, Miramare, Italy}
\vskip1.5cm
\vfill
\midinsert
\narrower
\noindent
{\bf Abstract.}
The path integral for the propagator is expanded into a perturbation
series, which can be exactly summed in the case of $\delta$-function
perturbations giving a closed expression for the (energy-dependent)
Green function. Making the strength of the $\delta$-function perturbation
infinite repulsive, produces a totally reflecting boundary, hence
giving a path integral solution in half-spaces in terms of the
corresponding Green function. The example of the Wood-Saxon potential
serves by an appropriate limiting procedure to obtain the Green function
for the step-potential and the finite potential-well in the half-space,
respectively.
\endinsert
\vskip1.5cm
\eject}
\pageno=1


\baselineskip=13.5pt
\glno=0                      
\line{\bf I Introduction \hfill}
\vglue 0.4cm
Although the technique of exactly computing Feynman path integrals
seems to reach a saturation point as far as generic Lagrangians with
smooth long range potentials are concerned, there are still a whole
range of problems which lack a systematic approach. Boundary problems
and piece-wise constant potentials belong to these classes of problems.
Steps forward to a solution of the latter have been done by e.g.\ Barut
and Duru$^{\BADU}$, for the former e.g.\ by Clark et al.$^{\CMS}$ and
Carreau et al.$^{\CAR}$. Barut and Duru obtained a formula for the
propagator for some piece-wise constant potentials via a canonical
transformation to Hamilton-Jacobi coordinates, however, left with
one (or more) additional integration(s) over momenta. In Refs.~[\CMS,
\CAR] boundary conditions could be implemented into the path integral
by means of cleverly chosen $\delta$-function perturbations in the
Lagrangian generating the corresponding boundary conditions. However,
it is often more appropriate to consider the (energy-dependent) Green
function $G(E)$ instead of the propagator $K(T)$. For instance, the
whole range of problems where a space-time transformation$^{\DK, \GRSb,
\KLE}$ must be performed demonstrates the convenient use of the Green
function $G(E)$.

In this paper I discuss boundary problems  with Dirichlet
boundary-conditions in path integral problems by explicitly stating
closed formul\ae\ for the Green functions. As we will see the
corresponding formul\ae\ can be derived by implementing a
$\delta$-function perturbation into the path integral, which leads to
an exactly summable perturbation expansion. Making the strength of the
$\delta$-function perturbation be infinite repulsive produces an
impenetrable boundary, i.e.\ Dirichlet boundary-conditios are generated.
It is possible to consider arbitrary one-dimensional potential problems
as long as the Green function for the problem without boundaries is
known. The specific example of the smooth-step, respectively the
Wood-Saxon potential, then gives by an appropriate limiting procedure
the path integral solution for the step-potential and finite potential
well in the half-space, respectively.


\vglue 0.6cm
\line{\bf II General method \hfill}
\vglue 0.4cm
The general method for the time-ordered perturbation expansion is quite
simple. We assume that we have a potential $W(x)=V(x)+\widetilde V(x)$
in the path integral and we suppose that $W$ is so complicated
that a  direct path integration is not possible. However, the path
integral corresponding to $V(x)$ is assumed to be known. We expand the
path integral containing $\widetilde V(x)$ in a perturbation expansion
about $V(x)$ in the following way. The initial kernel corresponding to
$V$ propagates in $\Delta t$-time unperturbed, then it interacts with
$\widetilde V$, propagates again in another $\Delta t$-time unperturbed,
a.s.o, up to the final state. This gives the series
expansion$^{\FH,\GODE}$ ($x\in\bbbr$)
\plus
$$\myalign
  &K(x'',x';T)
       =\int\limits_{x(t')=x'}^{x(t'')=x''}\CD x(t)\exp\Bigg\{\ih
  \int_{t'}^{t''}\bigg[{m\over2}\dot x^2-V(x)-\widetilde V(x)\bigg]dt\Bigg\}
  \\   &
  =K^{(V)}(x'',x';T)+\sum_{n=1}^\infty\bigg(-\ih\bigg)^n
  \left(\prod_{j=1}^n
  \int_{t'}^{t_{j+1}} dt_j\int_{-\infty}^\infty dx_j\right)
  \\   &\qquad\times K^{(V)}(x_1,x';t_1-t')
  \widetilde V(x_1)K^{(V)}(x_2,x_1;t_2-t_1)
  \times\dotsc
  \\   &\qquad\dots\times
  \widetilde V(x_{n-1})K^{(V)}(x_n,x_{n-1};t_n-t_{n-1})
  \times
  \widetilde V(x_n)K^{(V)}(x'',x_n;t''-t_n)\enspace.
  \tag\num\endalign$$
I have ordered time as $t'=t_0<t_1<t_2<\dots<
t_{n+1}=t''$ and paid attention to the fact that $K(t_j-t_{j-1})$ is
different from zero only if $t_j>t_{j-1}$.
We consider now an arbitrary one-dimensional potential $V(x)$ with an
additional $\delta$-function perturbation$^{\AGHH}$
\plus
$$W(x)=V(x)-\gamma\delta(x-a)\enspace.
  \tag\num$$
The path integral for this potential problem reads
\plus
$$\Kd(x'',x';T)
  =\int\limits_{x(t')=x'}^{x(t'')=x''}\CD x(t)\exp\Bigg\{\ih
   \int_{t'}^{t''}\bigg[{m\over2}\dot x^2-V(x)+\gamma\delta(x-a)
   \bigg]dt\Bigg\}\enspace.
  \tag\num$$
We have assumed that the path integral (Feynman kernel, respectively)
$K^{(V)}$ for the potential $V$ is known including its Green function
\plus
$$\aligned
  \GV(x'',x';E)
 &=\ih                    \int_0^\infty dT\,\e^{\i ET/\hbar}
   K^{(V)}(x'',x';T)
  \\
  K^{(V)}(x'',x';T)&=\int_{-\infty}^{\infty}{dE\over2\pi\i}
  \e^{-\i ET/\hbar}\GV(x'',x';E)\enspace.
  \endaligned
  \tag\num$$
\edef\numba{\num}%
Now, introducing the Green function $\Gd(E)$ of the perturbed system
similarly to (\numba), it is easy to sum up the emerging geometric
power series due to the convolution theorem of the
Fourier transformation, and we obtain$^{\GROh}$
\plus
$$\Gd(x'',x';E)
  =\GV(x'',x';E)-\dfrac{\GV(x'',a;E)\GV(a,x';E)}
              {\GV(a, a;E)-1/\gamma}\enspace,
  \tag\num$$
\edef\numbb{\num}%
where it is assumed that $\GV(a,a;E)$ actually exists. The
energy levels $E_n$ of the perturbed problem $W(x)$ are therefore
determined in a unique way by the denominator of $\Gd(E)$.
Radial problems, of course, can be discussed in a completely analogous
way. It is now straightforward to incorporate more than one
$\delta$-function perturbation. Considering two $\delta$-function
perturbations with strengths $\gamma_{1,2}$ located at $x=a_{1,2}$,
respectively, gives by a repetition of the previous procedure the Green
function $G^{(\delta_2)}(E)$ of the two-fold $\delta$-function perturbation:
\hfuzz=8pt
\plus$$G^{(\delta_2)}(x'',x';E)
  =\dsize\thickfrac{\left\vert\matrix
  \GV(x'',x';E)  &\GV(x'',a_1;E) &\GV(x'',a_2;E)    \\
  \GV(a_1,x';E)  &\GV(a_1,a_1;E)-1/\gamma_1
                                 &\GV(a_1,a_2;E)    \\
  \GV(a_2,x';E)  &\GV(a_2,a_1;E) &\GV(a_2,a_2;E)-1/\gamma_2
  \endmatrix\right\vert}{\left\vert\matrix
  \GV(a_1,a_1;E)-1/\gamma_1
                 &\GV(a_1,a_2;E)    \\
  \GV(a_2,a_1;E) &\GV(a_2,a_2;E)-1/\gamma_2
  \endmatrix\right\vert}\enspace.
  \tag\num$$
\edef\numbc{\num}%
\hfuzz=3pt
Equation (\numbc) has an obvious generalization to the problem of
$N$ $\delta$-function perturbations, which can be proven by induction.

In (\numbb) we consider now the limit $\gamma\to-\infty$ which has
the effect that an impenetrable wall appears$^{\CMS}$ at $x=a$.
We set $\lim_{\gamma\to-\infty}\Gd(E)\equiv G^{(Wall)}(E)$, i.e.\ we
obtain
\plus
$$\GWa(x'',x';E)
  =\GV(x'',x';E)-\dfrac{\GV(x'',a;E)\GV(a,x';E)}{\GV(a, a;E)}\enspace.
  \tag\num$$
Repeating the procedure for the two-fold $\delta$-function perturbation
Green function, we consider the limit $\lim_{\gamma_1,\gamma_2\to-\infty}$
$G^{(\delta_2)}(E)\equiv G^{(Box)}(E)$ and obtain
for the motion in the box $a<x<b$
\plus
$$G^{(Box)}(x'',x';E)
  =\dsize\thickfrac{\left\vert
  \matrix \GV(x'',x';E) &\GV(x'',b;E) &\GV(x'',a;E)  \\
          \GV(b,x';E)   &\GV(b,b;E)   &\GV(b,a;E)    \\
          \GV(a,x';E)   &\GV(a,b;E)   &\GV(a,a;E)
  \endmatrix\right\vert}{\left\vert\matrix
          \GV(b,b;E)   &\GV(b,a;E)    \\
          \GV(a,b;E)   &\GV(a,a;E)    \endmatrix\right\vert}
  \enspace.
  \tag\num$$
By the same method it is possible to consider motion constraint
by radial boxes and rings, respectively, with the corresponding
radial Green function $\GV_l(E)$ taken into account.


\vglue 0.6cm
\noindent
{\bf III The Wood-Saxon potential, the potential step, and the potential
         well}
\vglue 0.4cm
We consider the ``smooth step'' potential.
Its path integral solution is given by$^{\GROe, \KLEMUS}$
($b,R,V_0>0$ constants)
\plus
$$\myalign
  &\ih                   \int_0^\infty dT\,\e^{\i ET/\hbar}
  \int\limits_{x(t')=x'}^{x(t'')=x''}\CD x(t)
  \exp\left\{\ih\int_{t'}^{t''}\bigg[{m\over2}\dot x^2
      +{V_0\over 1+\e^{(x-b)/R}}\bigg]dt\right\}
  \\   &
  ={2mR\over\hbar^2}{\Gamma(m_1)\Gamma(m_1+1)\over
           \Gamma(m_1+m_2+1)\Gamma(m_1-m_2+1)}
  \\   &\qquad\times
  \bigg({1-\tanh{x'-b\over2R}\over2}\bigg)^{m_1-m_2\over2}
  \bigg({1+\tanh{x'-b\over2R}\over2}\bigg)^{m_1+m_2\over2}
  \\   &\qquad\times
  \bigg({1-\tanh{x''-b\over2R}\over2}\bigg)^{m_1-m_2\over2}
  \bigg({1+\tanh{x''-b\over2R}\over2}\bigg)^{m_1+m_2\over2}
  \\   &\qquad\times
  {_2}F_1\bigg(m_1,m_1+1;m_1-m_2+1;{1-\tanh{x_>-b\over2R}\over2}\bigg)
  \\   &\qquad\times
  {_2}F_1\bigg(m_1,m_1+1;m_1+m_2+1;{1+\tanh{x_<-b\over2R}\over2}\bigg)
  \enspace,
  \tag\num\endalign$$
with $x_{>,<}$ the larger/smaller of $x',x''$. Here denote
$m_{1,2}=\sqrt{2m}\,R\big(\sqrt{-E-V_0}\pm\sqrt{-E}\,\big)/\hbar$.
With a barrier at $x=a$, such that we consider motion in the
half-space $x>a$, we obtain that the Green function of the Wood-Saxon
potential is given by
\plus$$\myalign
   \ih&                   \int_0^\infty dT\,\e^{\i ET/\hbar}
  \int\limits_{x(t')=x'}^{x(t'')=x''}\CD_{(x>a)} x(t)
  \exp\left[\ih\int_{t'}^{t''}\bigg({m\over2}\dot x^2
      +{V_0\over 1+\e^{(x-b)/R}}\bigg)dt\right]
         \\   &
  ={2mR\over\hbar^2}{\Gamma(m_1)\Gamma(m_1+1)\over
           \Gamma(m_1+m_2+1)\Gamma(m_1-m_2+1)}
         \\   &\qquad\times
  \bigg({1-\tanh{x_<-b\over2R}\over2}\bigg)^{m_1-m_2\over2}
  \bigg({1+\tanh{x_<-b\over2R}\over2}\bigg)^{m_1+m_2\over2}
         \\   &\qquad\times
  \bigg({1-\tanh{x_>-b\over2R}\over2}\bigg)^{m_1-m_2\over2}
  \bigg({1+\tanh{x_>-b\over2R}\over2}\bigg)^{m_1+m_2\over2}
         \\   &\qquad\times\left\{
  {_2}F_1\bigg(m_1,m_1+1;m_1-m_2+1;{1-\tanh{x_>-b\over2R}\over2}\bigg)
  \right.\\   &\quad\qquad\times
  {_2}F_1\bigg(m_1,m_1+1;m_1+m_2+1;{1+\tanh{x_<-b\over2R}\over2}\bigg)
         \\   &\qquad\qquad-
  \dsize\thickfrac{{_2}F_1\bigg(m_1,m_1+1;m_1+m_2+1;
                     \dsize{1+\tanh{a-b\over2R}\over2}\bigg)}
  {{_2}F_1\bigg(m_1,m_1+1;m_1-m_2+1;
                             \dsize{1-\tanh{a-b\over2R}\over2}\bigg)}
         \\   &\qquad\qquad\qquad\times
  {_2}F_1\bigg(m_1,m_1+1;m_1-m_2+1;{1-\tanh{x'-b\over2R}\over2}\bigg)
         \\   &\qquad\qquad\qquad\times\left.
  {_2}F_1\bigg(m_1,m_1+1;m_1-m_2+1;{1-\tanh{x''-b\over2R}\over2}\bigg)
  \right\}\enspace,
  \tag\num\endalign$$
and the bound state energy levels
are determined by (with $0<\vert E_n\vert<V_0$)
\plus
$${_2}F_1\bigg(\beta_n-\i\lambda_n,\beta_n-\i\lambda_n+1;1+2\beta_n;
  {1-\tanh{a-b\over2R}\over2}\bigg)=0\enspace.
  \tag\num$$
Here denote $\beta^2=\beta^2(E)=-2mER^2/\hbar^2$,
$\lambda^2=\lambda^2(E)=2m(E+V_0)R^2/\hbar^2$, $\beta_n=\beta(E_n)$,
and $\lambda_n=\lambda(E_n)$.

In the limit $R\to0$ the smooth-step potential transforms into the
step-potential $V^{(PS)}(x)=[\Theta(x-b)-1]V_0$ with step-height $V_0$.
Using the transformation formula for the hypergeometric function
\plus
$$\multline
  {_2}F_1(\alpha,\beta;\gamma;z)
  ={\Gamma(\gamma)\Gamma(\gamma-\alpha-\beta)\over
    \Gamma(\gamma-\alpha)\Gamma(\gamma-\beta)}
   {_2}F_1(\alpha,\beta;\alpha+\beta-\gamma+1;1-z)
  \\
  +(1-z)^{\gamma-\alpha-\beta}
  {\Gamma(\gamma)\Gamma(\alpha+\beta-\gamma)\over
    \Gamma(\alpha)\Gamma(\beta)}
   {_2}F_1(\gamma-\alpha,\gamma-\beta;\gamma-\alpha-\beta+1;1-z)
  \endmultline
  \tag\num$$
we obtain for the Green function $G^{(PS)}(E)$ ($k\to\lambda/R=$
with $\sqrt{-2m(E+V_0)}/\hbar=-\i k$, $\chi\to\beta/R$)
\plus$$\myalign
   &G^{(PS)}(x'',x';E)
   \\
   &=\Theta(b-x')\Theta(b-x'')
   {1\over\hbar}\sqrt{-{m\over2(E+V_0)}}\,\e^{-\i k(x_<-b)}
   \bigg(\e^{\i k(x_>-b)}-{\chi+\i k\over\chi-\i k}\e^{-\i k(x_>-b)}\bigg)
   \\   &+\Theta(x'-b)\Theta(x''-b)
   {1\over\hbar}\sqrt{-{m\over2E}}\,
   \e^{-\chi(x_>-b)}
   \bigg(\e^{\chi(x_<-b)}+{\chi+\i k\over\chi-\i k}\e^{-\chi(x_<-b)}\bigg)
   \\   &+\Theta(x_>-b)\Theta(b-x_<)
   {1\over\hbar}{\sqrt{2m}\over\sqrt{-E}\,+\sqrt{-E-V_0}}
   \e^{-\i k(x_<-b)} \e^{-\chi(x_>-b)}
   \enspace,
   \tag\num\endalign$$
[alternatively we can write $e^{2\i\arctan(k/\chi)}=(\chi+\i k)/
\chi-\i k)$]. The continuity of the Green function at the location
of the step at $x=b$ is easily checked and one obtains
$$\allowdisplaybreaks\alignat 3
  G^{(PS)}(x_>,b;E)
  &={2m\over\hbar^2(\chi-\i k)}\e^{-\chi(x_>-b)}
  &\qquad
  &(x_>>b,\,x_<=b)
  \tag\num\\   \global\plus
  G^{(PS)}(x_<,b;E)
  &={2m\over\hbar^2(\chi-\i k)}\e^{-\i k(x_<-b)}
  &\qquad
  &(x_<<b,\,x_>=b)\enspace.
  \tag\num\endalignat$$

Considering now a totally reflecting barrier at $x=a<b$ we obtain
for the potential well (PW) in the half-space $x>a$ the Green function
\hfuzz=5pt
\plus$$\myalign
   &G^{(PW)}(x'',x';E)
   \\   &=\Theta(b-x')\Theta(b-x'')
   {1\over\hbar}\sqrt{-{m\over2(E+V_0)}}\Bigg\{\e^{-\i k(x_<-b)}
   \bigg(\e^{\i k(x_>-b)}-{\chi+\i k\over\chi-\i k}\e^{-\i k(x_>-b)}\bigg)
   \\   &\qquad
   - \bigg(\e^{2\i k(a-b)}-{\chi+\i k\over\chi-\i k}\bigg)^{-1}
   \\
   &\qquad\qquad\times
   \bigg(\e^{\i k(x''-b)}-{\chi+\i k\over\chi-\i k}\e^{-\i k(x''-b)}\bigg)
   \bigg(\e^{\i k(x'-b)}-{\chi+\i k\over\chi-\i k}\e^{-\i k(x'-b)}\bigg)
   \Bigg\}
   \\   &+\Theta(x'-b)\Theta(x''-b)\Bigg\{
   {1\over\hbar}\sqrt{-{m\over2E}}\,\e^{-\chi(x_>-b)}
   \bigg(\e^{\chi(x_<-b)}+{\chi+\i k\over\chi-\i k}\e^{-\chi(x_<-b)}\bigg)
   \\   &\qquad
   -{2\over\hbar}{\sqrt{-2m(E+V_0)}\over(\sqrt{-E}\,+\sqrt{-E-V_0}\,)^2}
   \bigg(\e^{2\i k(a-b)}-{\chi+\i k\over\chi-\i k}\bigg)^{-1}
   \e^{-\chi(x''-b)-\chi(x'-b)}\Bigg\}
   \\   &+\Theta(x_>-b)\Theta(b-x_<)
   {1\over\hbar}{\sqrt{2m}\over\sqrt{-E}\,+\sqrt{-E-V_0}}
   \Bigg\{\e^{-\i k(x_<-b)} \e^{-\chi(x_>-b)}
   \\   &\qquad
   -\bigg(\e^{2\i k(a-b)}-{\chi+\i k\over\chi-\i k}\bigg)^{-1}
   \bigg(\e^{\i k(x_<-b)}-{\chi+\i k\over\chi-\i k}\e^{-\i k(x_<-b)}\bigg)
   \e^{-\chi(x_>-b)}\Bigg\}\enspace.
   \tag\num\endalign$$
\hfuzz=3pt
Again, the continuity at $x=b$ and the boundary conditions of the Green
function are easily checked. The bound state energy
levels are determined by the poles of $G^{(PW)}$ yielding to the
well-known result$^{\FLU}$
\plus$$
  {k\over\chi}=-\tan k(b-a)\enspace.
  \tag\num$$


\vglue 0.6cm
\line{\bf IV Discussion \hfill}
\vglue 0.4cm
In this paper I have presented a perturbation expansion approach to path
integral problems with Dirichlet boundary-conditions. The use of a
perturbation expansion was necessary because the construction of the
Feynman kernel of a Dirichlet problem by means of the ``mirror''
principle fails generally because the entire kernel does not have the
required reflection symmetry. I obtained closed formul\ae\ of the
corresponding Green functions for arbitrary systems put into half-spaces,
boxes, radial boxes and rings, respectively. Particularly, in the case
of radial boxes we can consider the corresponding motion inside a radial
box, respectively, outside a hard sphere. Of course, numerous examples
could serve to demonstrate the power of the presented formalism, say,
the linear potential in the half-space, the radial harmonic oscillator
including a $1/r^2$-term inside a radial box, motion under the influence
of a Aharonov-Bohm solenoid outside a hard disc located at the origin, and
many more.

As examples of the technique I chose the cases of the smooth-step
potential, respectively the Wood-Saxon potential, which gave the path
integral solution in terms of the corresponding Green functions for the
step-potential and the finite potential-well in the half-space,
respectively. These two examples are of considerable importance, because
the treatment of piece-wise constant potentials in the path integral
have been very rudimentary up to now. Hence, two examples of an entire
new class of quantum mechanical problems are added to the list of
exactly solvable path integrals$^{\GRSb}$. The present treatment also
has in contrast to Ref.~[\BADU] the advantage of stating explicitly
simple quantization conditions for bound state solutions.

What remains is a thorough discussion of, say, Neumann
boundary-conditions in the path integral along the lines presented
here for Dirichlet problems. This topic, however, will be addressed
elsewhere.


\vglue 0.6cm
\line{\bf References \hfill}
\baselineskip=10pt
\vglue 0.4cm
\eightpoint
\eightrm
\def\refno{\item}
\refno{$^{\BADU}$}
A.O.Barut and I.H.Duru:
Path Integration Via Hamilton-Jacobi Coordinates and Application to
Potential Barriers;
{\it Phys.Rev.}\ {\bf A 38} (1988) 5906
\refno{$^{\CMS}$}
T.E.Clark, R.Menikoff and D.H.Sharp:
Quantum Mechanics on the Half-Line Using Path Integrals;
{\it Phys.Rev.}\ {\bf D 22} (1980) 3012
\refno{$^{\CAR}$}
M.Carreau:
The Functional Integral for a Free Particle on a Half-Plane;
{\it J.Math.Phys.}\ {\bf 33} (1992) 4139;
\newline
M.Carreau, E.Farhi and S.Gutmann:
Functional Integral for a Free Particle in a Box;
{\it Phys.Rev.}\ {\bf D 42} (1990) 1194
\refno{$^{\DK}$}
I.H.Duru and H.Kleinert:
Solution of the Path Integral for the H-Atom;
{\it Phys.Lett.}\ {\bf B 84} (1979) 185;
\newline
Quantum Mechanics of H-Atoms from Path Integrals;
{\it Fortschr.Phys.}\ {\bf 30} (1982) 401
\refno{$^{\GRSb}$}
C.Grosche and F.Steiner:
Path Integrals on Curved Manifolds;
{\it Zeitschr.Phys.}\ {\bf C 36} (1987) 699;
\newline
Classification of Solvable Feynman Path Integrals;
{\it DESY preprint} DESY 92-189, to appear in the {\it Proceedings of
the ``Fourth International Conference on Path Integrals from $meV$ to
$MeV$'', May 1992, Tutzing, Germany} (World Scientific, Singapore);
\newline
Table of Feynman Path Integrals;
to appear in: {\it Springer Tracts in Modern Physics}
\refno{$^{\KLE}$}
H.Kleinert:
Path Integrals in Quantum Mechanics, Statistics and Polymer Physics
(World Scientific, Singapore, 1990)
\refno{$^{\FH}$}
R.P.Feynman and A.Hibbs: Quantum Mechanics and Path Integrals
(McGraw Hill, New York, 1965)
\refno{$^{\GODE}$}
M.J.Goovaerts, A.Babceno, and J.T.Devreese,
A New Expansion Method in the Feynman Path Integral Formalism:
Application to a One-Dimensional Delta-Function Potential;
{\it J.Math.Phys.}\ {\bf 14} (1973) 554;
\newline
M.J.Goovaerts and J.T.Devreese,
Analytic Treatment of the Coulomb Potential in the Path Integral
Formalism by Exact Summation of a Perturbation Expansion;
{\it J.Math.Phys.}\ {\bf 13} (1972) 1070;
and Erratum: {\it J.Math.Phys.}\ {\bf 14} (1973) 153
\refno{$^{\AGHH}$}
S.Albeverio, F.Gesztesy, R.J.H\o egh-Krohn and H.Holden:
Solvable Models in Quantum Mechanics
(Springer Verlag, Berlin, 1988)
\refno{$^{\GROh}$}
C.Grosche:
Path Integrals for Potential Problems With $\delta$-Function
Perturbation;
{\it J.Phys.A: Math.Gen.}\ {\bf 23} (1990) 5205
\refno{$^{\GROe}$}
C.Grosche:
Path Integral Solution of a Class of Potentials Related to the
P\"oschl-Teller Potential;
{\it J.Phys.A: Math.Gen.}\ {\bf 22} (1989) 5073
\refno{$^{\KLEMUS}$}
H.Kleinert and I.Mustapic:
Summing the Spectral Representations of P\"oschl-Teller and
Rosen-Morse Fixed-Energy Amplitudes;
{\it J.Math.Phys.}\ {\bf 33} (1992) 643
\refno{$^{\FLU}$}
S.Fl\"ugge:
Practical Quantum Mechanics, Vol.I (Springer-Verlag, Berlin, 1977)


\enddocument